# A Framework For DevaNagari Script-Based CAPTCHA

Sushma Yalamanchili [1] and Kameswara Rao[2]

[1]Research Scholar, Department of Computer Science & Engineering
Acharya Nagarjuna University, Andhra Pradesh, India.
sushma_yalamanchili@yahoo.co.in

[2]Department of Computer Science, P.G.Centre, P.B.Siddhartha College, Vijayawada.
Kamesh.manchiraju@gmail.com

*ABSTRACT*

*Human Interactive Proofs (HIPs) are automatic reverse Turing tests designed to distinguish between various groups of users. Completely Automatic Public Turing test to tell Computers and Humans Apart (CAPTCHA) is a HIP system that distinguish between humans and malicious computer programs. Many CAPTCHAs have been proposed in the literature that text-graphical based, audio-based, puzzle-based and mathematical questions-based. The design and implementation of CAPTCHAs fall in the realm of Artificial Intelligence. We aim to utilize CAPTCHAs as a tool to improve the security of Internet based applications. In this paper we present a framework for a text-based CAPTCHA based on Devanagari script which can exploit the difference in the reading proficiency between humans and computer programs. Our selection of Devanagari script-based CAPTCHA is based on the fact that it is used by a large number of Indian languages including Hindi which is the third most spoken language. There is potential for an exponential rise in the applications that are likely to be developed in that script thereby making it easy to secure Indian language based applications.*

## 1. INTRODUCTION

Human Interactive Proofs (HIPs) [1] focus on automation tests that virtually all humans can pass but current computer programs fail [2]. Completely Automated Public Turing test to tell Computers and Humans Apart (CAPTCHA) was an acronym that was coined in 2000. It is a type of challenge-response test that only a human completes successfully. In the simplest form of a CAPTCHA, an image consisting of digits and letters sufficiently distorted is presented and the user is required to input the characters that are displayed. Other forms of CAPTCHAs are based on text-graphics, audio, hand-writing and puzzles. CAPTCHAs have been widely used as a security measure to restrict access from Robots or Bots. CAPTCHAs are based on Artificial Intelligence (AI) problems that cannot be solved by current computer programs or Bots but are easily solvable by humans. A client who provides a correct response to a challenge is presumed to be a human; otherwise a Bot.

## 2. CURRENT RESEARCH

In this section, we review the current research on CAPTCHAs and Optical Character Recognition (OCR) efforts for the Devanagari script. Websites use CAPTCHAs as a security measurement to distinguish human users from Bots. While CAPTCHAs have been developed based on pure text, images, audio and video, text CAPTCHAs are almost exclusively used in real applications. In a text CAPTCHA, characters are deliberately distorted and connected to prevent recognition by Bots. Security of an existing text CAPTCHA is enhanced by systematically adding noise and distortion, and arranging characters more tightly [3, 4]. Usability is always an important issue in designing a CAPTCHA [5]. Examples of text-based CAPTCHAs include the Gimpy method [6], the Baffletext method [7], Handwritten CAPTCHA [8], the PayPal method [9], Using Dynamic Visual Patterns [10], the Hotmail Method [11] and Pessimal Print method [12]. Successful text CAPTCHAs used by Microsoft, Yahoo and Google use techniques that are resistant to segmentation [13, 14, and 15] attacks by using random acrs, connected random lines and crowding characters.

Figure 1 shows an example of a HIP challenge presented when registering for a MSN Hotmail account. After displaying the HIP, the user is asked to type the eight characters included within it, namely, X29JTUN3.

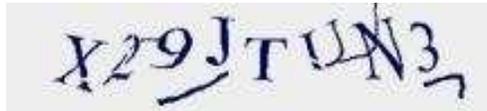

Figure 1 Microsoft Hotmail Text CAPTCHA

A few of the many different HIP styles that can be produced by manipulating hardness parameters are illustrated below (Figure 2).

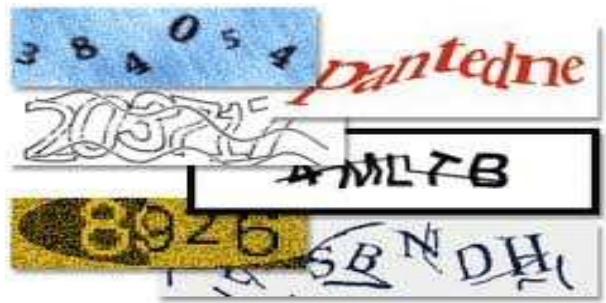

Figure 2 Hardness Parameter variations in HIPs

Image-based CAPTCHAs [16] require users to identify labeled images or rotated images. They evince a larger gap between human users and Bots because of the poor ability of Bots in obtaining features of images. In Image Recognition CAPTCHAs, the user is provided with a small set of images to name or distinguish or identify anomalies. In implicit CAPTCHAs [17], the user does not have to read or type anything and just makes simple clicks on hot spots. Drawing captcha [18] generates numerous dots on a screen with noisy background, some of which are diverse from the others. The user has to connect them to each other. Apart from the visual-CAPTCHAs, there exist a number of audio CAPTCHAs [19, 20, and 21] where the user must recognize and type the word that is played as a sound. Sound-based CAPTCHAs exploit a border range of human abilities which are mainly based on the auditory perception of human ability to identify words or letters in a sound clip after being distorted and with additive background noise. Their emergence has greatly aided vision impaired people. A typical sound-based CAPTCHA is reCAPTCHA [22] proposed by Carnegie Mellon University and later acquired by Google. Video-based CAPTCHAs also substitute a brief video display of characters for the usual letters. In video-based CAPTCHAs [23], users will be prompted to view a challenge video and then appropriately annotate (or tag) it. The challenges will be graded by exact matching of user response with a database of ground truth tags for the video.

Multilingual CAPTCHAs [24, 25] have been proposed in literature. In [24] the user can select his/her native language and subsequent screens are displayed in that language using translation.

The user is expected to select an image corresponding to the specified topic. For instance, a user is required to select an image of a cat from among pictures of a car, a human, scenery or a flower. The picture of the cat may also be rotated which requires a Bot to have knowledge of the shape of the object that needs to be selected as well as the complexity involved on account of the rotation. Design for a CAPTCHA based on dictionary words in Persian/Arabic is discussed with refinements [25]. The premise of the Persian/Arabic Baffletext CAPTCHA is that adding a background to a random assortment of Persian/Arabic characters interferes with OCR recognition of the text string due to the presence of diacritical dots and signs that are used in that script. The characteristics of Persian/Arabic script that are utilized in the CAPTCHA design include word lengthening, lack of spacing between words, different forms of the same character depending on placement in a word, overlap of characters, different letter sizes and separation of vowels from letters.

Researchers enjoyed moderate success as part of computer vision research in breaking existing text-based CAPTCHAs and image-based CAPTCHAs [26, 27]. Image recognition is considered to be a much harder problem than text recognition. OCR deals with automatic recognition of different characters in a document image leading to clear and unambiguous recognition, analysis and understanding of the document content. OCR system segments text zone into text lines, text lines into words, and words into characters. These characters are then recognized. The task of recognition can be broadly separated into two categories: machine printed data and the handwritten data. Researchers have investigated OCR for Devanagari script [28, 29]. These systems use statistical, syntactic and/or heuristic-based methods. Pre-processing, Segmentation, Feature Extraction, Recognition and Post processing are major stages in OCR [30, 31]. The success or failure of an OCR system depends on segmentation and feature extraction stages. Efforts have been made to improve the performance of OCR applied to Devanagari script [32, 33]. Still a lot of research is needed to tackle the challenges in identifying specific characters of Devanagari script. The characteristics of Devanagari script include major issues in OCR segmentation and feature extraction stages. The major factors that affect the OCR process in Devanagari script include noise, intra-word and inter-word touching, shape of compound character formation, change of font and style (Glyph Variation), multiple skewness, broken characters, etc. Many works on OCR efforts for Devanagari script have been reported [34, 35, 36, and 37].

## 3. MOTIVATION

Popular web sites are subject to brute force attacks by programs or computers known as Robots or Bots. They can be used to break user accounts or submit an unlimited number of service requests such as email account creation, web connection requests and serving as shopping agents. Such activities often lead to abuse of privilege causing the server to exhaust its resources or worse cause it to shut down. Bots can be eliminated by introducing CAPTCHA at service request time to prevent Denial of Service attacks. Introducing CAPTCHA in the authentication scheme prevents automated brute force attack.

We propose a framework for text-based CAPTCHA that is based on Devanagari script that is the written form for several Indian languages including Hindi, Sanskrit, Kashmiri, Bihari, Bhojpuri, Marati and Nepali. As the population that uses these languages is extremely large, we have selected the Devanagari script for use in this CAPTCHA. DevaCAPTCHA can be used to enhance the security of Devanagari script-based Indian language content based web applications. Some applications that are typical in this scenario are online and collaborative authoring, online tutoring, email and social networking portals in Indian languages. These users are typically involved in content generation and access of applications in Devanagari-script based Indian languages. They already have the necessary keyboard emulation or are using custom keyboards that facilitate keyboard input in that script so that aspect is taken care of.

# 4. DEVANAGARI SCRIPT

The Devanagari script is a Brahmi-derived writing system used originally to write Sanskrit. It is used in India and Nepal to write many languages including Sanskrit, Hindi, Marathi, Nepali, Konkoni etc. Devanagari is the script used by more than 400 million people on the globe. Devanagari has 11 vowels and 33 simple consonants. Besides the consonants and the vowels, other constituent symbols in Devanagari are set of vowel modifiers called *matra* (placed to the left, right, above, or at the bottom of a character or *conjunct*), pure-consonant (also called half-letters) which when combined with other consonants yield conjuncts. A horizontal line called *shirorekha* runs through the entire span of work. Devanagari word is written into the three strips namely: a core strip, a top strip, and a bottom strip as shown in Figure 3. The core strip and top strip are differentiated by the header, while the lower modifier is attached to the core character.

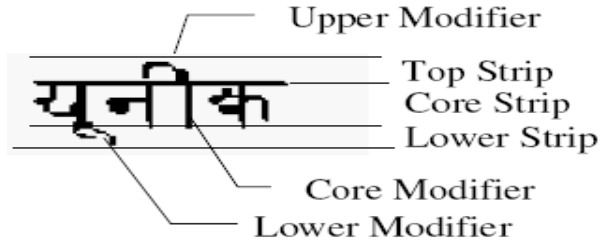

Figure 3 Character strips in Devanagari script

OCR for Devanagari script becomes even more difficult when compound character and modifier characteristics are combined in 'noisy' situations. The Devanagari script character set is depicted in Figures 4 and 5.

(a) Vowels अ आ इ ई उ ऊ ऋ ए ऐ ओ औ
(b) Modifier Symbols corresponding to the vowels (the modifier symbol has also been attached to the consonant क to indicate its placing
का कि की कु कू कृ के कै को कौ
(c) Consonants क ख ग घ ङ च छ ज झ ञ ट ठ ड ढ ण त थ द ध न प फ ब भ म य र ल व श ष स ह
(d) Pure Consonants
(e) Some Conjuncts formed by Pure Consonants modifiers when combined with character य
क्य ख्य घ्य च्य ज्य त्य थ्य ध्य न्य प्य भ्य म्य य्य ल्य व्य

Figure 4 Devanagari script character set

० १ २ ३ ४ ५ ६ ७ ८ ९ १०
0 1 2 3 4 5 6 7 8 9 10

Figure 5 Devanagari script Numerals

# 5. PROPOSED FRAMEWORK

In this work, we propose a framework for DevaCAPTCHA (Figure 6) for administering a Devanagari script-based text CAPTCHA to assist in securing Indian-language web-based applications. The objective of a DevaCAPTCHA framework is to differentiate a human from a Bot. This can be achieved by testing the ability of a user to recognize the Devanagari characters.

The key components of the proposed DevaCAPTCHA framework are

1. DevaDB: A sufficiently large database of devanagari script samples (either in text or handwritten form)
2. Query Generator: A mechanism to query the database and obtain a random sample subject to the design
3. Obfuscator: A module that takes the random sample from the database that distorts and adds noise to it
4. DevaGUI: User challenge-response interface
5. Match Response: A determination of whether the user has submitted the accurate response for the challenge posed.

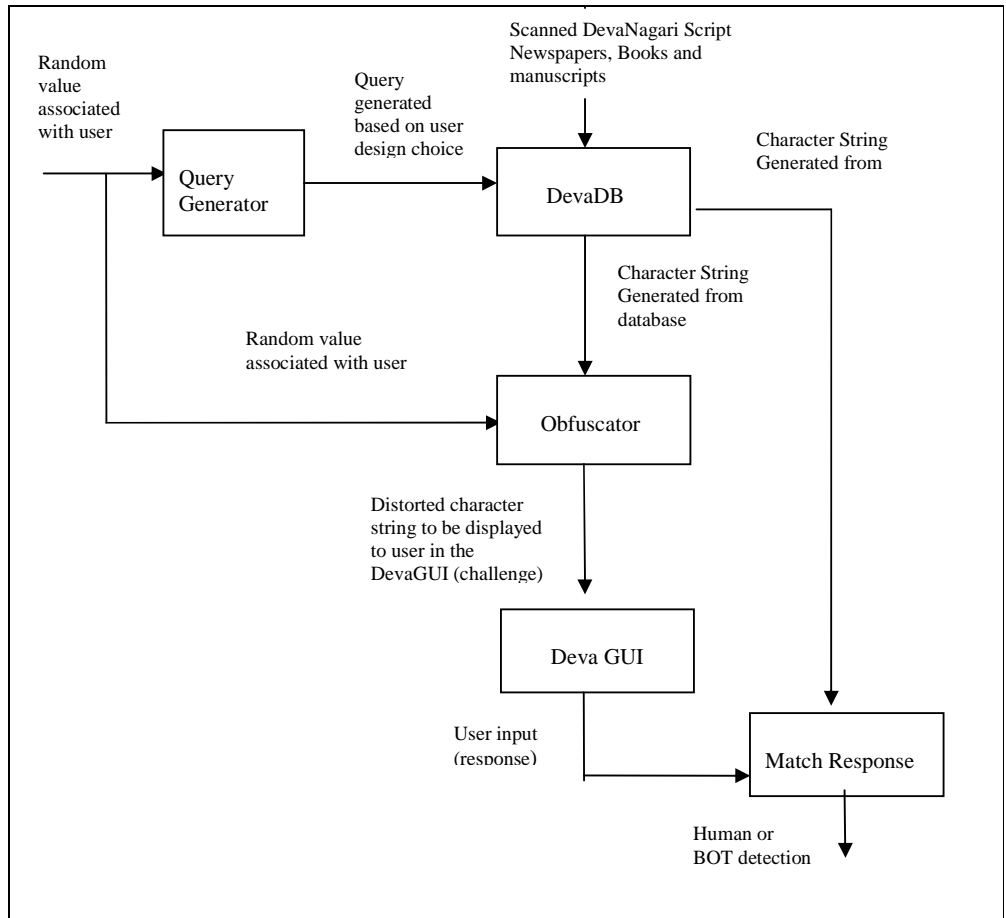

Figure 6 DevaCAPTCHA framework

## 5.1 DevaDB

DevaDB is a collection of books, newspapers and manuscripts in Devanagari script. The database is required to be online and can be shared by many web applications. The database can exclusively contain pages in Devanagari script based languages or may be multi-lingual allowing support for many other scripts/languages. The database population process is a one-time major effort with periodic minor updates or addition. The contents are stored as pages which are uniquely identified as bit objects. The pages are indexed and a list of words with associated pages is generated. Indexing is also a one-time major effort with re-indexing performed only on updation. This can be used for randomly selecting a pre-existing word from the database. Similar procedures can be adopted to generate phrases from the pages. In case a random word is required, a sub-module of the DevaDB module is responsible for generation of a random word character string with the suitable string length determined by the query generator. The Query Generator module prepares a query which is fed to the Database module which outputs a random word, a word or a phrase as a Devanagari script-based character string. Thus DevaDB can be used to serve digital content as well as provide the source for generating the challenge.

Work is already in progress partially funded by the Government of India and other agencies for creation of Digital Libraries in many research centers and in many Indian languages not limited to Devanagari script-based languages. The aim of these projects is to digitize Indian language books and manuscripts for better preservation and easy access. In these projects, the initial phase involves scanning of old newspapers, manuscripts and books and storing the content as images. In other projects, the Indian language content is converted into text files using OCR technology. The latter approach is suitable for DevaDB. Additionally, a web crawler may be developed and used in the Database module that will populate the database with copyright-expired content available on the web. Unicode is used to support the different types of fonts that are available for Devanagari script.

### 5.2 Query Generator

The Query Generator module simply generates a query to DevaDB that will determine whether the text string is a valid word or not, whether the length of the string is fixed in length or variable. Other variations can be if the text string is a phrase or a word. A random number that is associated with the user determines the parameters associated with the text string that is to be generated by the database. The minimum and maximum length of the word, whether we select an existing word from the database or whether we generate a random assortment of characters are design choices or can be user specified in this module without compromising the security of the Turing test.

### 5.3 Obfuscator

The Obfuscator module is the crux of DevaCAPTCHA and uses specific characteristics of Devanagari script to deceive the adversary. In Devanagari script, all the individual characters are joined by a head line called "Shiro Rekha". The obfuscator may remove the headline and adds noise to the image using patterns like mosaic, arcs/jaws, vertically overlapped on the script. The Obfuscator further misleads machine by using different fonts of different sizes with variable character spacing. Sample images that are intended to be used in DevaCAPTCHA are shown below (Figure 7). Segmentation resistant methods will be employed in its design.

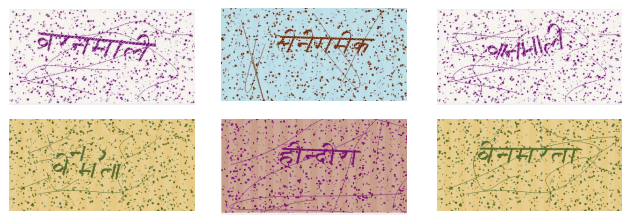

Figure 7 Visualized DevaCAPTCHA challenge strings

The Obfuscator module operates using a Transformation function that is applied to the character string generated by DevaDB. The Transformation function consists of N (in the range of 50-100) parameters each of which represents a particular type of alteration that can be applied to the character string. The alterations may be at the individual character component level or at the level of the entire character string. Periodically a subset of m (m < N) parameters are selected randomly that will be used for this time period. Thus it is ensured that Bots are unaware of which subset of N parameters are being currently used. The random input associated with each user will determine the values of each of the m parameters.

For instance, parameter 1 may indicate font variation, parameter 2 may indicate font size variation, parameter 3 may indicate variable character separation (overlapping characters, shadow characters, joint characters at different points), parameter 4 may indicate skew, parameter 5 may indicate noise (heavy typesetting, frayed paper/manuscript, faded print, additive and subtractive noise), parameter 6 may indicate inclusion of a background that interferes with the symbols, parameter 7 may indicate removal of shirorekha, parameter 8 may indicate inclusion of a curve passing through the character string, parameter 9 may indicate multiple consonant conjuncts, parameter 10 may indicate character level distortions (stretching, compression), parameter 11 may indicate representation of the character string along a curve, etc.

The extent of distortions applied is determined by the degree of robustness and usability that is required. Robustness refers to the ability of a CAPTCHA to prevent Bots from successfully deciphering the character string. Usability refers to easy solvability by humans. A highly robust CAPTCHA successfully deflects Bot attacks but may also present greater difficulty for humans to solve. Thus robustness and usability aspects have to be carefully balanced. The distortions applied should be difficult for Bots to solve but very easily solved by humans.

### 5.4 DevaGUI

In DevaCAPTCHA, we display the substantially noisy and distorted image containing the chosen text string or word in the DevaGUI (Figure 8) image display area. A text input box is provided where the user can type in the characters in sequence as they appear in the distorted image. A submit button is to be pressed to signal completion of input.

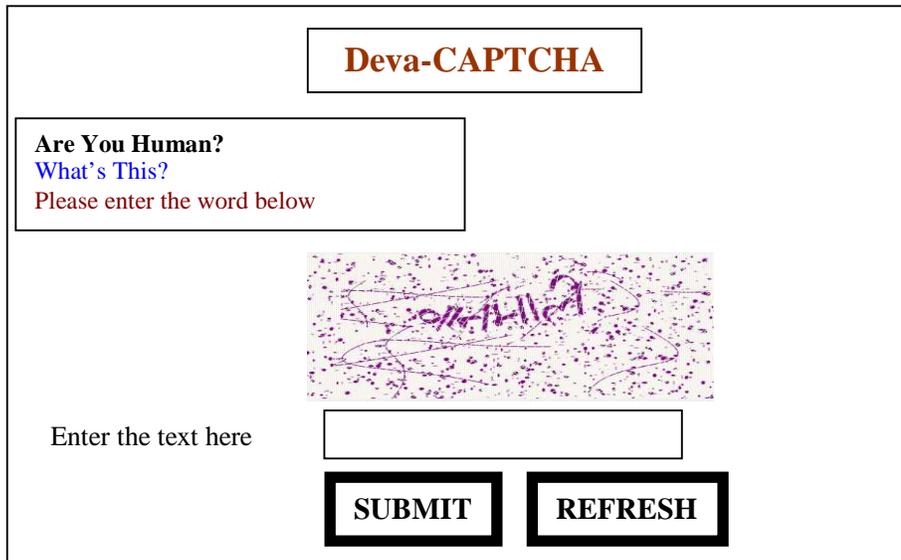

Figure 8 DevaGUI

A timer can be used to cause expiration of a CAPTCHA image so that users are given a specified amount of time to solve the CAPTCHA while Bots have limited time to break the CAPTCHA. Upon expiration of the current CAPTCHA image, the user can request a new CAPTCHA string to be displayed by pressing the Refresh button. In case the user is unable to solve the current CAPTCHA, the Refresh button may be used to request a new string. When the user is able to successfully solve the string, he is permitted to move onto the next phase of authentication or request a service from the web application.

**5.5 Match Response**

The user typed in text from DevaGUI should match the pre-obfuscated string generated by the Database in response to the DevaDB query. If it does, the user is deemed to be a human and may be permitted to authenticate or request for web services.

## 6. CONCLUSION

We have seen that by distinguishing between humans and computers, CAPTCHAs offer protection against automated attacks on systems and applications. The criterion for success of a text-based CAPTCHA are its robustness and usability. We have outlined techniques in the Obfuscator module to generate the challenge that render DevaCAPTCHA robust as it is resistant to Bot attacks that employ OCR technology.

DevaCAPTCHA is highly usable as it is easy for humans to successfully provide the response. Since we are using words from books and newspapers or an assortment of characters that are non-words, it is not difficult for humans to visually perceive these characters despite the distortions and noise due to their superior visual capabilities and cognitive abilities to make connections with words that they have encountered in some context. Distorted images containing random strings are still easy for humans to read while computers spend endless time processing information. The implementation of DevaCAPTCHA and the participation in OCR testing efforts related to Indian language scripts is to be taken up as future research work. Handwriting recognition and testing for Devanagari script is another future research activity.

**Authors**

Sushma Yalamanchili completed her undergraduate studies in Computer Science at BITS, Pilani, India and graduate studies at Michigan State University, USA. She has been teaching graduate students since 1998. Her areas of interest are Network & Information Security, Human Interaction Proof, Pattern Recognition and Image Processing.

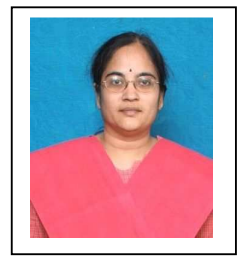

M. Kameswara Rao has MCA from Acharya Nagarjuna University and M.Phil from Bharatidasan University. He is currently pursuing M.Tech in Acharya Nagarjuna University. He has taught graduate students in Computer Science for ten years. His areas of interest are Network Security, Pattern Recognition and Image Processing.

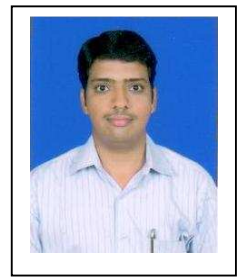